# Effect of Joule heating in current-driven domain wall motion

A. Yamaguchi, H. Tanigawa[1], T. Ono[1], S. Nasu, K. Miyake[2], K. Mibu[3], and T. Shinjo[4]

Graduate School of Engineering Science, Osaka University, 1-3 Machikaneyama-cho, Toyonaka, 560-8531, Japan

[1]Institute for Chemical Research, Kyoto University, Gokasho, Uji, 611-0011, Japan

[2]Graduate School of Engineering, Tohoku University, 5, Aza-Aoba, Aramaki, Aoba-ku, Sendai, 980-8579, Japan

[3]Research Center for Low-Temperature and Materials Sciences, Kyoto University, Gokasho, Uji, 611-0011, Japan

[4]International Institute for Advanced Studies, 9-3 Kizugawadai, Kizu-cho, Soraku-gun, 619-0225, Japan

# ABSTRACT

It was found that high current density needed for the current-driven domain wall motion results in the Joule heating of the sample. The sample temperature, when the current-driven domain wall motion occurred, was estimated by measuring the sample resistance during the application of a pulsed-current. The sample temperature was 750 K for the threshold current density of $6.7 \times 10^{11}$ A/m² in a 10 nm-thick $Ni_{81}Fe_{19}$ wire with a width of 240 nm. The temperature was raised to 830 K for the current density of $7.5 \times 10^{11}$ A/m², which is very close to the Curie temperature of bulk $Ni_{81}Fe_{19}$. When the current density exceeded $7.5 \times 10^{11}$ A/m², an appearance of a multi-domain structure in the wire was observed by magnetic force microscopy, suggesting that the sample temperature exceeded the Curie temperature.

The current-driven domain wall (DW) motion due to the spin transfer effect has been confirmed by experiments on magnetic thin films and magnetic wires [1-9]. This effect makes it possible to switch the magnetic configuration without an external magnetic field, and is of importance for future spintronic devices. However, the current density required is relatively high, of the order of $10^{11}$ A/m$^2$, and it is necessary to investigate the effect of the Joule heating for practical applications.

We estimated the sample temperature by comparing the sample resistance during the application of a pulsed-current with the independently measured temperature dependence of the sample resistance. It was found that the sample temperature got close to the Curie temperature of Ni$_{81}$Fe$_{19}$, when the current-driven DW motion occurred.

Special L-shaped magnetic wires of 10 nm-thick Ni$_{81}$Fe$_{19}$ were fabricated onto thermally oxidized Si substrates by means of an *e*-beam lithography and a lift-off method as schematically illustrated in Fig. 1 [7]. The width of the wire is 240 nm. One end of the L-shaped magnetic wire is connected to a diamond-shaped pad which acts as a DW injector [10], and the other end is sharply pointed to prevent the nucleation of a DW from this end [11]. The wire has four electrodes made of nonmagnetic material, 20 nm-thick Cu, for electrical transport measurements.

The resistance during the application of a pulsed-current was measured with the circuit schematically shown by the inset in Fig. 2. To determine the amplitude of the pulsed-current flowing through the circuit, $I$, the reference resistance, $R$, was connected in series with the sample. The sample resistance during the application of the pulsed-current, $R_{pulse}$, is determined as

$$R_{pulse} = \frac{V_2}{I} = \frac{V_2}{V_1} \cdot R.$$

Here, $V_1$ and $V_2$ are the voltages induced in the reference resistance and in the sample, respectively. $V_1$ and $V_2$ were measured by a digital storage oscilloscope. $R_{pulse}$ was measured for various pulsed-current densities with a duration of 5 μs.

The standard temperature dependence of the sample resistance was measured independently in the temperature range between 200 K and 400 K with small constant current.

The time dependences of $V_1$ and $V_2$ are superimposed in Fig. 2. Here, a pulsed-voltage of 1 V with a duration of 5 μs was applied to the circuit. $V_1$

and $V_2$ reached to the constant values within 1 μs, suggesting that the resistance of the sample was almost constant during the application of the pulsed-current.

Figure 3(a) shows $R_{pulse}$ as a function of current density. $R_{pulse}$ increases with the current density, and $R_{pulse}$ for the threshold current density was almost twice the resistance at room temperature. This result suggested the high current density caused considerable Joule heating of the sample.

The temperature dependence of the resistance measured with constant current of 1 μA is shown in Fig. 3(b). Above 300 K, the resistance was almost proportional to the temperature. We crudely extrapolated the linear temperature dependence of the resistance between 300 K and 400 K into higher temperatures. The dashed line is the extrapolated temperature dependence of the resistance.

By comparing the result of Fig. 3(a) with that of Fig. 3(b), we found that the sample temperature was about 750 K for the threshold current density of $6.7 \times 10^{11}$ A/m². The temperature was raised to 830 K for the current density of $7.5 \times 10^{11}$ A/m², which is very close to the Curie temperature of bulk $Ni_{81}Fe_{19}$ (850 K).

We also checked the magnetic domain structure of the sample before and after applying a pulsed-current. After the observation with magnetic force microscopy (MFM) shown in Fig. 4(a), which confirmed the existence of a single tail-to-tail DW, a pulsed-current with a duration of 5 μs was applied through the wire in the absence of a magnetic field. In the current density range from $6.7 \times 10^{11}$ A/m² to $7.5 \times 10^{11}$ A/m², the current-driven DW motion has been experimentally confirmed [7,12].

When the current density exceeded $7.5 \times 10^{11}$ A/m², an appearance of the multi-domain structure in the wire was observed as shown in Fig. 4(b). A multi-domain structure was observed even for the sample without a DW by the application of the pulsed-current exceeded $7.5 \times 10^{11}$ A/m². These experimental results suggested the sample temperature exceeded the Curie temperature of $Ni_{81}Fe_{19}$ when the current density exceeded about $7.5 \times 10^{11}$ A/m², which is in agreement with the estimation of the sample temperature from the sample resistance during the application of the pulsed-current.

We have shown that the high current density required for the current-driven domain wall motion resulted in the considerable Joule heating and that the sample temperature got close to the Curie temperature when the current-driven DW motion occurred. Thus, we should take into account thermal effects, such as decrease in magnetic moment, excitation of spin-waves, and thermal depinning of a DW for the understanding of the current-driven domain wall motion.

The present work was partly supported by the Ministry of Education, Culture, Sports, Science and Technology of Japan through Special Coordination Funds for Promoting Science and Technology (Nanospintronics Design and Realization, NDR) and by the 21st Century COE Program by Japan Society for the Promotion of Science.

**FIGURE CAPTION**

Figure 1 Schematic illustration of the top view of the sample.

Figure 2 Output voltages induced in the reference resistance R (left axis) and in the sample $R_{pulse}$ (right axis) as a function of time. Inserted illustration is the circuit for the measurement.

Figure 3 (a) Resistance of the sample during the application of the pulsed-current as a function of the current density. (b) Standard temperature dependence of the sample resistance with constant current 1 μA.

Figure 4 (a) MFM image after the introduction of a DW. The DW is imaged as a dark contrast, which corresponds to the stray field from negative magnetic charge. (b) MFM image after the application of a pulsed-current with the current density of $7.5 \times 10^{11}$ A/m². Multi-domain structure appears in the wire.

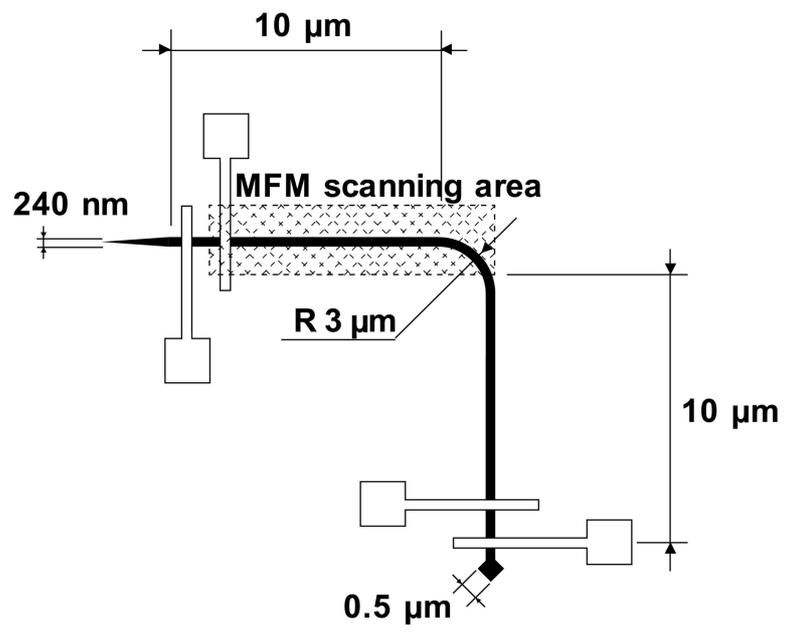

Fig. 1

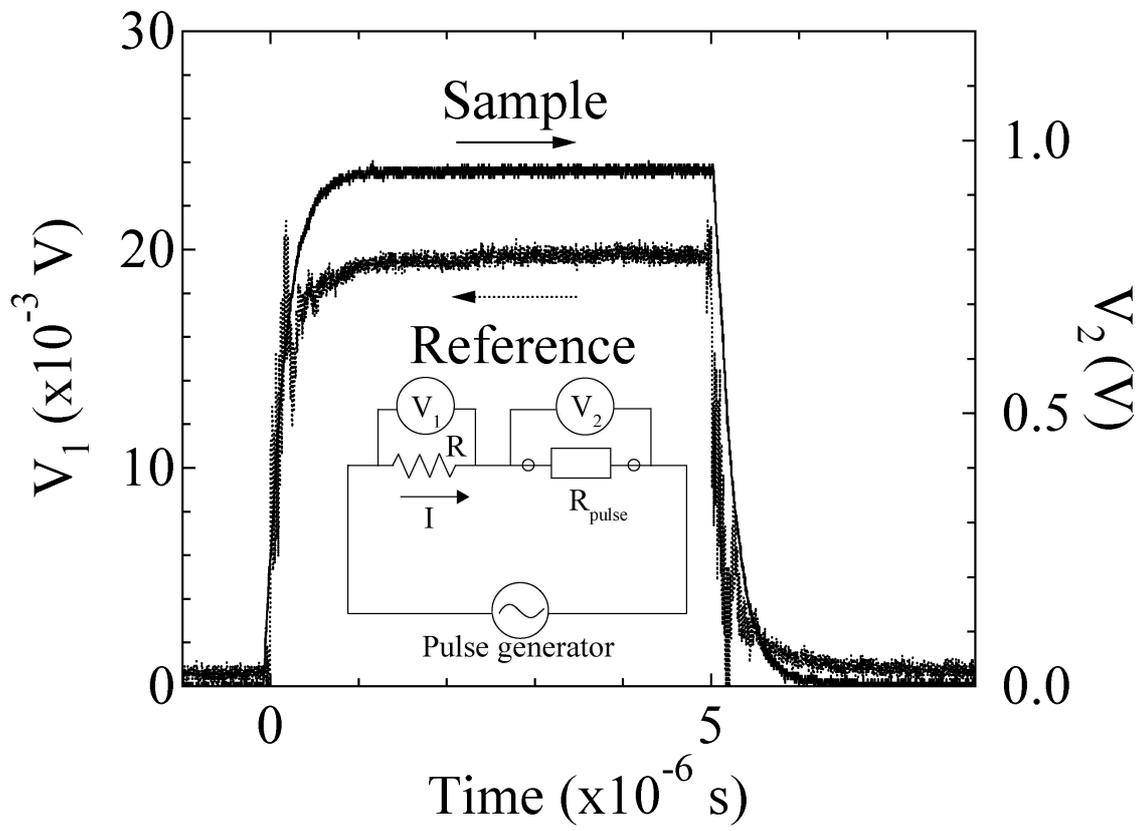

Fig. 2

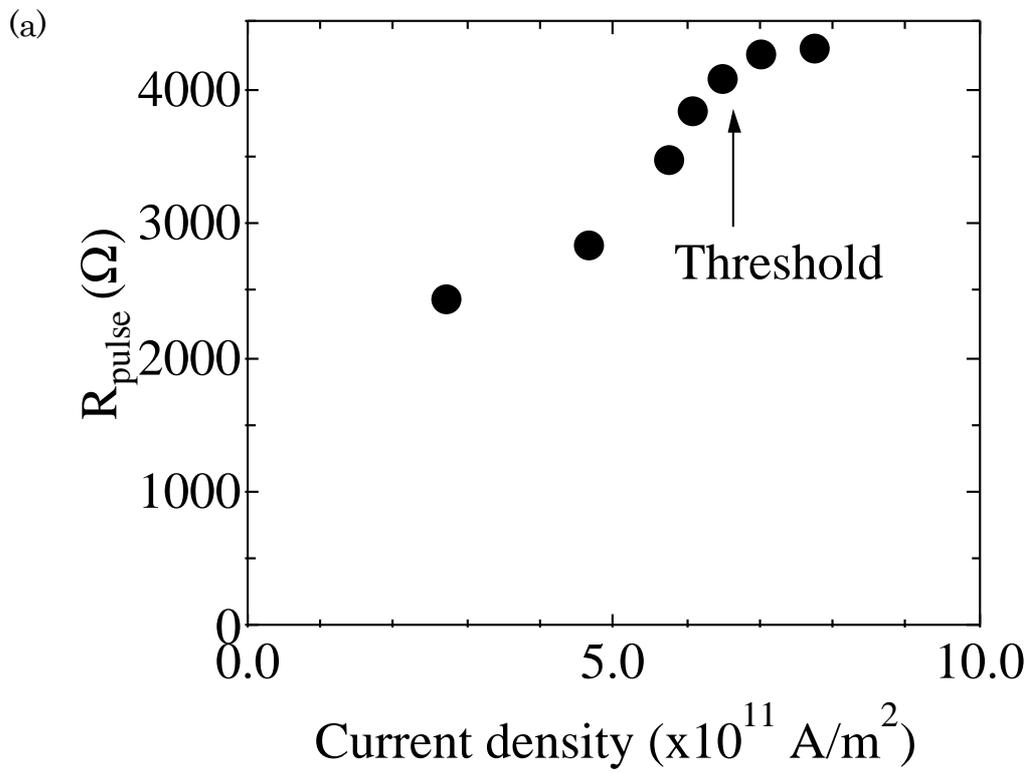

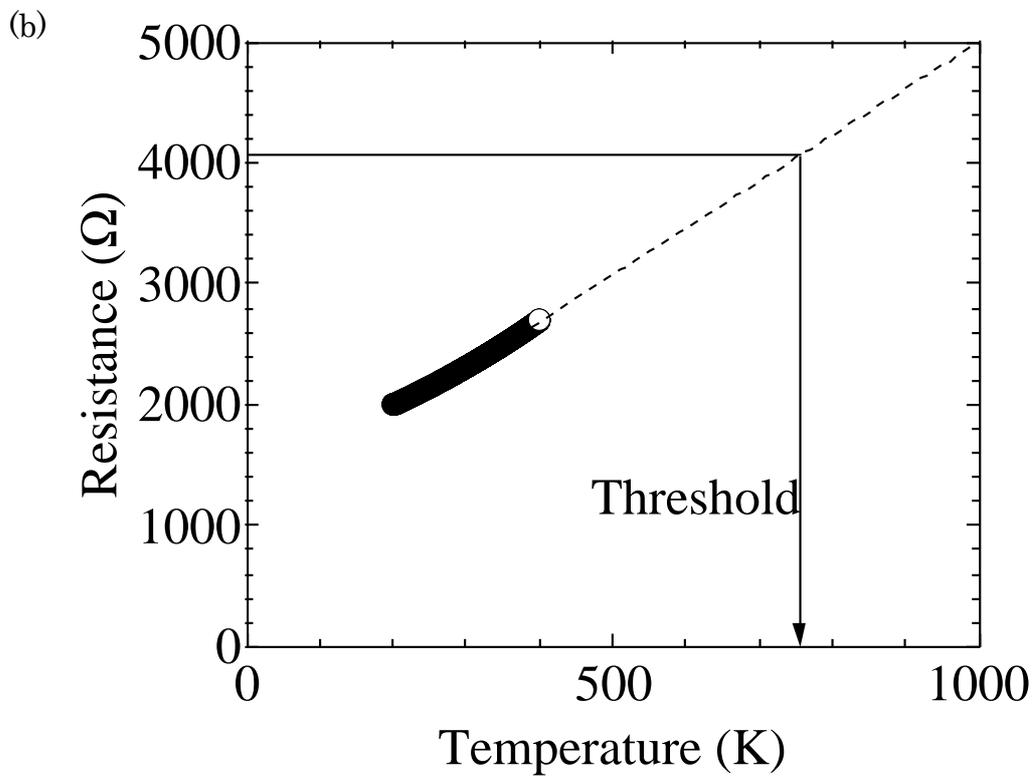

Fig. 3

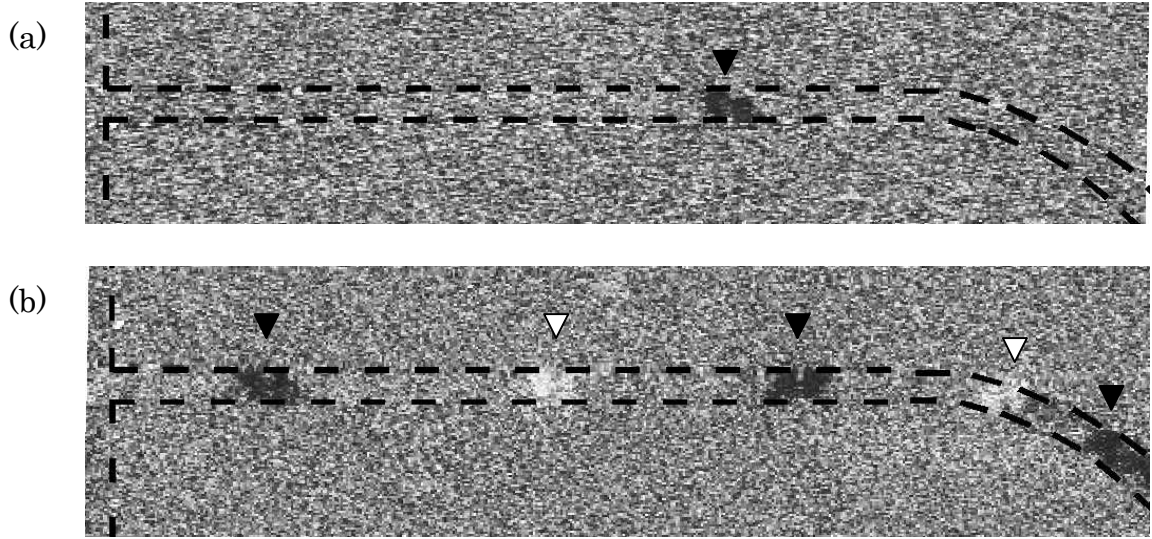

Fig. 4